\documentclass[english,aps,pra,showpacs,superscriptaddress,twocolumn]{revtex4}
\usepackage[T1]{fontenc}
\usepackage{amsmath,graphicx,amssymb,babel,dsfont,color}

\def\bbm[#1]{\mbox{\boldmath $#1$}}
\renewcommand{\Re}{\text{Re}}
\renewcommand{\Im}{\text{Im}}
\newcommand{\Tr}{\text{Tr}}

\newcommand{\rd}{\text{d}}
\newcommand{\kb}{k_{\rm B}}
\newcommand{\ri}{\text{i}}
\newcommand{\re}{{\rm e}}

\begin{document}

\title{Fluctuational-electrodynamic theory and\\dynamics of heat transfer in multiple dipolar systems}

\author{Riccardo Messina}\email{riccardo.messina@institutoptique.fr}
\affiliation{Laboratoire Charles Fabry, UMR 8501, Institut d'Optique, CNRS, Universit\'{e} Paris-Sud 11, 2, Avenue Augustin Fresnel, 91127 Palaiseau Cedex, France.}
\author{Maria Tschikin}
\affiliation{Institut f\"{u}r Physik, Carl von Ossietzky Universit\"{a}t, D-26111 Oldenburg, Germany.}
\author{Svend-Age Biehs}
\affiliation{Institut f\"{u}r Physik, Carl von Ossietzky Universit\"{a}t, D-26111 Oldenburg, Germany.}
\author{Philippe Ben-Abdallah}\email{pba@institutoptique.fr}
\affiliation{Laboratoire Charles Fabry, UMR 8501, Institut d'Optique, CNRS, Universit\'{e} Paris-Sud 11, 2, Avenue Augustin Fresnel, 91127 Palaiseau Cedex, France.}

\date{\today}

\begin{abstract}
A general fluctuational-electrodynamic theory is developed to investigate radiative heat exchanges between objects which are assumed small compared with their thermal wavelength (dipolar approximation) in $N$-body systems immersed in a thermal bath. This theoretical framework is applied to study the dynamic of heating/cooling of three-body systems. We show that many-body interactions allow to tailor the temperature field distribution and to drastically change the time scale of thermal relaxation processes.
\end{abstract}


\maketitle

\section{Introduction}

The absence of thermal equilibrium is at the origin of an energy exchange between bodies having different temperatures mediated by the electromagnetic field. This radiative heat transfer was first described by Planck's theory in the far field. Stefan-Boltzmann law, which is valid only when the distance $d$ between the bodies is large compared to the thermal wavelength $\lambda_\text{th}=\hbar c/\kb T$ ($\hbar$ is Planck's reduced constant, $c$ the speed of light in vacuum, $\kb$ Boltzmann's constant, and $T$ the temperature), which is of the order of some microns at ambient temperature~\cite{Planck91}. It was later shown that in the near-field regime, i.e.\ when $d\ll \lambda_\text{th}$, the heat transfer can surpass its far-field counterpart by several orders of magnitude. This effect was first predicted in the pioneering work of Polder and van Hove~\cite{PolderPRB71} using the approach based on fluctuational-electrodynamic theory developed by Rytov~\cite{Rytov89}. According to this approach, each body is described by a distribution of fluctuating currents, whose statistical properties are connected through the fluctuation-dissipation theorem to the temperatures and dielectric properties of the bodies.

It has been shown that the near-field amplification is mainly due to the tunneling of evanescent photons, which do not participate to the exchange in the far field~\cite{LoomisPRB94,PendryJPhysCondensMatter99,VolokitinPRB01,VolokitinPRB04,JoulainSurfSciRep05,VolokitinRevModPhys07}. This amplification is even more remarkable if the bodies support surface resonances, such as plasmons for metals or phonon-polaritons for polar materials: in this case the heat transfer is almost monochromatic around the surface-resonance frequency~\cite{MuletApplPhysLett01,MuletMicroscaleThermophysEng02}. The experimental confirmation of the near-field enhancement of heat transfer is now well-established in sphere-plane and plane-plane geometries~\cite{KittelPRL05,HuApplPhysLett08,NarayanaswamyPRB08,RousseauNaturePhoton09,ShenNanoLetters09,KralikRevSciInstrum11,OttensPRL11,vanZwolPRL12a,vanZwolPRL12b,KralikPRL12}. Moreover, the study of radiative heat transfer can be relevant for several applications, going from thermophotovoltaic~\cite{DiMatteoApplPhysLett01,NarayanaswamyApplPhysLett03,LarocheJApplPhys06,BasuIntJEnergyRes07,FrancoeurApplPhysLett08,BasuJApplPhys09} or solar thermal energy conversion~\cite{SwnasonScience09,Chen11} to heat-assisted data storage~\cite{SrituravanichNanoLett04}.

During the last three years several theories have been developed to describe heat transfer at any separation distance between bodies with arbitrary geometries and dielectric properties. Having in common the use of fluctuation-dissipation theorem, these approaches differ in the technique employed: scattering matrices~\cite{BimontePRA09,MessinaEurophysLett11,MessinaPRA11}, Green's functions~\cite{KrugerPRL11,KrugerPRB12}, time-domain calculations~\cite{RodriguezPRL11}, boundary-element methods~\cite{McCauleyPRB12} and fluctuating surface currents~\cite{RodriguezPRB12,RodriguezarXiv13}. Although some of these theories~\cite{KrugerPRL11,RodriguezarXiv13} allow in principle to treat the case of an arbitrary number of bodies, the numerical applications have been performed only in the case of two bodies having several different geometries.

Recently, a step forward has been performed by investigating heat transfer in three-body systems. The case of three dipoles~\cite{BenAbdallahPRL11} and three parallel planar slabs~\cite{ZhengNanoscale11,MessinaPRL12} have been described in detail. These results have shown new promising ways, using many-body interactions, to produce interesting effects, such as the inhibition or amplification of heat flux, by exploiting the intrinsically non-additive behavior of radiative heat transfer.

The dynamics of heat transfer in the near field has also been recently addressed. In Ref.~\cite{TschikinEurPhysJB12} the cooling/heating of a nanoparticle immersed in a thermal bath close to a planar surface is considered, discussing how it depends on the particle-surface distance. Very recently, Yannopapas and Vitanov~\cite{YannopapasPRL13} have extended this study to a collection of nanoparticles and outlined the possibility of thermal control by means of an external laser source. They investigated the possibilities of controlling the temperature distribution within a collection of metallic nano-particles by means of an external coherent laser field. However the interaction of nanoparticles with their surroundings is taken into account using an heuristic approach based on the introduction of an average absorption cross section. Further, a quantum description of the heat transfer dynamics for two plasmonic nanoparticles was developed in Ref.~\cite{BiehsAgarwal2013}. Finally, it must be mentioned that other authors have considered both theoretically and experimentally the nano-scale control of the time-independent temperature profile for a system made of metallic nano-structures~\cite{BaffouOptExpress09,BaffouPRB10,BaffouLaserPhotonicsRev13}. The nano-scale control (both time-dependent and time-independent) of the temperature distribution has proved to be crucial for several applications, such as heat-assisted nano-chemistry~\cite{CaoNanoLett07,AdlemanNanoLett09,ChristopherNatureChem11} or thermotherapy for medical applications, in particular in the context of cancer treatment~\cite{FalkIntJHyperthermia01,vanderZeeAnnOncol02,GuNanotoday07,HirschProcNatlAcadSciUSA03,ONealCancerLetter04}.

In this work we introduce a general theory to describe the time-dependent heat flux and temperature distribution for an arbitrary number $N$ of particles, described using the dipolar approximation, immersed in a thermal bath at constant temperature. Using a purely fluctuational-electrodynamic approach, we deduce the expression of the power absorbed by each particle, isolating the contributions coming from each other particle and from the bath. Differently from~\cite{YannopapasPRL13}, we provide a general derivation of all the contributions to the energy exchanges (both between particles and with the external thermal bath) based on the fluctuation-dissipation theorem. In the case of three nano-spheres, we use this knowledge to study the dynamics of the temperatures when one of the particles is initially heated up, without any external energy source during the time evolution. We discuss the influence of the geometrical configuration as well as that of the coupling of surface resonances.

This paper is organized as follows. In Sec.~\ref{SecPhysSyst} we introduce the physical system and the main equations describing the time evolution and the power absorbed by each particle. In Sec.~\ref{SecTotalpE} we find a closed-form analytical expression for the total dipole moment associated to each particle and for the electric field at the particle positions. Section \ref{SecAP} contains the derivation of the power absorbed by each particle, identifying the contributions coming from each other particle and from the bath. In Sec.~\ref{SecNumerics} we provide some numerical applications: we study the dynamics of the temperatures in a three-particle system, discussing the influence of geometry and surface resonances; we also discuss how the distribution of particles can be used to produce different time-independent temperature profiles. Finally, in Sec.~\ref{SecConcl} we draw our conclusions.

\section{Physical system and energy balance}\label{SecPhysSyst}

We consider a discrete set of $N$ objects at different temperatures $T_i$ centered at positions $\mathbf{r}_i$ inside a thermal bath (a free bosonic field) which is maintained at temperature $T_\text{b}$. While $T_\text{b}$ is assumed to be fixed, the $N$ temperatures $T_i$ can vary in time. We suppose that the sizes of the objects are small compared with the smallest thermal wavelength $\lambda_{T_i}=c\hbar/(k_\text{B}T_i)$ so that all individual objects can be modeled as simple radiating electrical dipoles $\mathbf{p}_i$ and magnetic dipoles $\mathbf{m}_i$. Here we limit our discussion to non-magnetic materials (i.e.\ $\mathbf{m}_i=\mathbf{0}$). We assume that the time scale produced by radiative heat exchanges is large compared to the phonon thermalization time in each object (typically of the order of some picoseconds for a nanoparticle). Under this hypothesis, whose validity will be discussed in Sec.~\ref{SecNumerics}, it is meaningful to define a temperature $T_i(t)$ for each particle as a function of time. Assuming also no phase and mass change of materials, the time evolution of the $N$ temperatures $T_i$ is governed by the following energy equations $(i = 1, \ldots, N)$
\begin{equation}
  \label{TempEv0}
  \rho_i C_i V_i \frac{\rd T_i(t)}{\rd t}=-\int_{S_i} \langle \mathbf{\Pi}(\mathbf{r},t)\rangle\cdot \rd\mathbf{S}_i,
\end{equation}
where the LHS is the time variation of the internal energy of object $i$, $\rho_i$, $C_i$, and $V_i$ representing its mass density, heat capacity and volume, respectively. The RHS determines the energy flux across the oriented surface $S_i$ enclosing the particle with a dipole moment $\mathbf{p}_{0i}(\mathbf{r},t)=\mathbf{p}_i(t)\delta(\mathbf{r}-\mathbf{r}_i)$ by integrating the Poynting vector $\mathbf{\Pi}$ over $S_i$. In expression \eqref{TempEv0} the brackets represent the ensemble average over all the statistical realizations. In the context of a quantum treatment of field and matter, this average (and all the quantum averages from now on) has to be intended as a symmetrized average $\langle\mathcal{A}\mathcal{B}\rangle_\text{sym}=(\langle\mathcal{A}\mathcal{B}\rangle+\langle\mathcal{B}\mathcal{A}\rangle)/2$. At local thermal equilibrium we have, according to the Poynting theorem,
\begin{equation}
  \label{EqPoynting}
  \nabla\cdot\mathbf{\Pi}(\mathbf{r},t)=-\mathbf{j}_i(\mathbf{r},t)\cdot\mathbf{E}(\mathbf{r},t),
\end{equation}
where $\mathbf{j}_i\cdot\mathbf{E}$ is the power dissipated by Ohmic losses in the volume $V_i$, $\mathbf{j}_i=\frac{\rd \mathbf{p}_{0i}}{\rd t}$ being the local electric current density and $\mathbf{E}$ the local electric field at position $\mathbf{r}$. Thus, by transforming the surface integral appearing in Eq.~\eqref{TempEv0} into a volume integral we cast Eq.~\eqref{TempEv0} into the form
\begin{equation}
  \label{TempEv}
  \rho_i C_i V_i \frac{\rd T_i(t)}{\rd t} = \wp_i^\text{(abs)}(t,T_1,\dots,T_N,T_\text{b}),
\end{equation}
where the power $\wp_i^\text{(abs)}$ absorbed by the dipole $i$ is given by
\begin{equation}
   \label{AbsP0}
   \begin{split}
      \wp_i^\text{(abs)}(t,T_1,\dots,T_N,T_\text{b})&=\int_{V_i}\!\langle\,\mathbf{j}_i(\mathbf{r},t)\cdot\mathbf{E}(\mathbf{r},t)\rangle\, \rd V_i\\
                                                    &=\biggl\langle\frac{\rd\mathbf{p}_i(t)}{\rd t}\cdot\mathbf{E}(\mathbf{r}_i,t)\biggr\rangle.
   \end{split}
\end{equation}
In order to calculate the absorbed power, we deduce in the next section an explicit expression of the electric field and dipole moment.

\section{Total dipole moment and field}\label{SecTotalpE}

We start by decomposing the local field $\mathbf{E}(\mathbf{r})$ into its incident part (which correspond to the bosonic field $\mathbf{E}^{\text{(b)}}$ of bath without scatterers) and its induced part $\mathbf{E}^{\text{(ind)}}$ as (we will from now on work in the frequency domain and omit the frequency dependence when only one frequency is concerned)
\begin{equation}
  \label{FieldDec0}
  \mathbf{E}(\mathbf{r})=\mathbf{E}^{\text{(b)}}(\mathbf{r})+\mathbf{E}^{\text{(ind)}}(\mathbf{r}).
\end{equation}
We then express the latter with respect to the electric dyadic Green tensor $\mathds{G}^{(0)}=\mathds{G}^\text{EE}$ as a function of all dipolar moments
\begin{equation}
  \label{FieldDec}
  \mathbf{E}(\mathbf{r}) = \mathbf{E}^{\text{(b)}}(\mathbf{r})+\frac{k^2}{\varepsilon_0}\sum_i\mathds{G}^{(0)}(\mathbf{r},\mathbf{r}_i)\mathbf{p}_i,
\end{equation}
where
\begin{equation}
  \label{DefG0}
  \begin{split}
    \mathds{G}^{(0)}(\mathbf{r},\mathbf{r}') &= \frac{\exp(\ri k\rho)}{4\pi\rho}\Bigl[\Bigl(1+\frac{\ri k\rho-1}{k^2\rho^2}\Bigr)\mathds{1}\\
                                             &\qquad +\frac{3(1-\ri k\rho)-k^2\rho^2}{k^2\rho^2}\widehat{\bbm[\rho]}\otimes\widehat{\bbm[\rho]}\Bigr]
  \end{split}
\end{equation}
is the dyadic Green tensor in free space, $k=\omega/c$, $\widehat{\mathbf{r}}=\mathbf{r}/r$, $\bbm[\rho]=\mathbf{r}'-\mathbf{r}$ and $\rho=|\bbm[\rho]|$. We now decompose each dipole moment $\mathbf{p}_i$ into
\begin{equation}
   \label{TotDipole}
   \mathbf{p}_i = \mathbf{p}_i^\text{(fl)}+\mathbf{p}_i^\text{(ind)},
\end{equation}
where $\mathbf{p}_i^\text{(fl)}$ and $\mathbf{p}_i^\text{(ind)}$ denote its fluctuating and induced parts respectively. For the induced part $\mathbf{p}_i^\text{(ind)}$ we use the discrete-dipole approximation~\cite{DraineAstrophysJ88,LakhtakiaJModPhysC92,Novotny06}, according to which $\mathbf{p}_i^\text{(ind)}$ is expressed as a function of the exciting field, i.e.\ the local field at $\mathbf{r}=\mathbf{r}_i$ except the contribution of the dipole $i$, as
\begin{equation}
  \label{IndDip}
  \mathbf{p}_i^\text{(ind)} = \varepsilon_0\alpha_i\Bigl(\mathbf{E}_i^{\text{(b)}}+\frac{k^2}{\varepsilon_0}\sum_{j\neq i}\mathds{G}^{(0)}_{ij}\mathbf{p}_j\Bigr)
\end{equation}
where $\alpha_i$ represents the (frequency-dependent) polarizability of dipole $i$ (assumed for simplicity isotropic) and we have introduced the notation $\mathbf{E}_i=\mathbf{E}(\mathbf{r}_i)$ and set $\mathds{G}_{ij}^{(0)}=\mathds{G}^{(0)}(\mathbf{r}_i,\mathbf{r}_j)$. Using Eqs.~\eqref{TotDipole} and \eqref{IndDip} we obtain the following equality, written in matrix form
\begin{equation}\label{EqMatrixDip}\begin{pmatrix}\mathbf{p}_1\\\vdots\\\mathbf{p}_N\end{pmatrix}=
\mathds{T}^{-1}\begin{pmatrix}\mathbf{p}_1^\text{(fl)}\\\vdots\\\mathbf{p}_N^\text{(fl)}\end{pmatrix}+
\mathds{T}^{-1}\mathds{A}\begin{pmatrix}\mathbf{E}_1^{\text{(b)}}\\\vdots\\\mathbf{E}_N^{\text{(b)}}\end{pmatrix}.\end{equation}
$\mathds{A}$ and $\mathds{T}$ are $3N\times3N$ block matrices defined in terms of the $(i,j)$ $N\times N$ sub-matrices ($i,j=1,\dots,N$)
\begin{equation}
  \label{DefTA}
  \mathds{T}_{ij} = \delta_{ij}\mathds{1}-(1-\delta_{ij})k^2\alpha_i\mathds{G}_{ij}^{(0)},\quad\mathds{A}_{ij}=\delta_{ij}\varepsilon_0\alpha_i\mathds{1}.
\end{equation}
For the local field we have, using Eqs.~\eqref{FieldDec} and \eqref{EqMatrixDip}
\begin{equation}
  \begin{pmatrix} \mathbf{E}_1\\\vdots\\\mathbf{E}_N \end{pmatrix} =
     \mathds{D}\mathds{T}^{-1}\begin{pmatrix}\mathbf{p}_1^\text{(fl)}\\\vdots\\\mathbf{p}_N^\text{(fl)}
  \end{pmatrix}
    + \bigl(\mathds{1}+\mathds{D}\mathds{T}^{-1}\mathds{A})
      \begin{pmatrix}\mathbf{E}_1^{\text{(b)}}\\\vdots\\\mathbf{E}_N^{\text{(b)}}\end{pmatrix},
\end{equation}
with
\begin{equation}
  \mathds{D}_{ij}=\frac{k^2}{\varepsilon_0}\mathds{G}_{ij}^{(0)}.
\end{equation}
It is easy to prove that
\begin{equation}
  \mathds{D} = -\mathds{A}^{-1}\mathds{T}+\mathds{B},\quad
  \mathds{B}_{ij} = \delta_{ij}\Bigl(\frac{1}{\varepsilon_0\alpha_i}\mathds{1}+\frac{k^2}{\varepsilon_0}\mathds{G}_{ii}^{(0)}\Bigr),
\end{equation}
and then
\begin{equation}\label{EqMatrixE}\begin{pmatrix}\mathbf{E}_1\\\vdots\\\mathbf{E}_N\end{pmatrix}=
\bigl(\mathds{B}\mathds{T}^{-1}-\mathds{A}^{-1}\bigr)\begin{pmatrix}\mathbf{p}_1^\text{(fl)}\\\vdots\\\mathbf{p}_N^\text{(fl)}\end{pmatrix}+
\mathds{B}\mathds{T}^{-1}\mathds{A}\begin{pmatrix}\mathbf{E}_1^{\text{(b)}}\\\vdots\\\mathbf{E}_N^{\text{(b)}}\end{pmatrix}.\end{equation}
Equations \eqref{EqMatrixDip} and \eqref{EqMatrixE} contain the expression of the total dipole moment and local electric field as a function of the fluctuating dipole moments and field of the bath. These expressions will be used to deduce, in the next section, the total power absorbed by each dipole.

\section{Exchanged powers}\label{SecAP}

Starting from Eq.~\eqref{AbsP0} we obtain
\begin{equation}
  \label{AbsP}
  \begin{split}
    \wp_i^\text{(abs)}&(t,T_1,\dots,T_N,T_\text{b}) \\
    & = -\ri \int_0^{+\infty} \!\! \frac{\rd\omega}{2\pi}\, \omega \int_0^{+\infty} \!\! \frac{\rd\omega'}{2\pi}\,\\
    & \qquad \times \Bigl[\langle\mathbf{p}_i(\omega)\cdot\mathbf{E}_i^\dag(\omega')\rangle \re^{-\ri(\omega-\omega')t} \\
    & \qquad\qquad - \langle\mathbf{p}_i^\dag(\omega)\cdot\mathbf{E}_i(\omega')\rangle \re^{\ri(\omega-\omega')t}\Bigr]\\
    & = 2\int_0^{+\infty}\!\!\frac{\rd\omega}{2\pi}\,\omega\int_0^{+\infty}\!\!\frac{\rd\omega'}{2\pi}\, \\
    &\qquad\times \Im\Bigl[\langle\mathbf{p}_i(\omega)\cdot\mathbf{E}_i^\dag(\omega')\rangle \re^{-\ri(\omega-\omega')t}\Bigr],
  \end{split}
\end{equation}
where we consider only positive frequencies and we use the convention $f(t)=2\,\Re\Bigl[\int_0^{+\infty}\!\!\frac{\rd\omega}{2\pi}f(\omega)\re^{-\ri\omega t}\Bigr]$ for the time Fourier transform. We now assume the general linear relations
\begin{equation}
  \begin{split}
    \begin{pmatrix}\mathbf{p}_1\\\vdots\\\mathbf{p}_N\end{pmatrix}&=
    \mathds{M}\begin{pmatrix}\mathbf{p}_1^\text{(fl)}\\\vdots\\\mathbf{p}_N^\text{(fl)}\end{pmatrix}+
    \mathds{N}\begin{pmatrix}\mathbf{E}_1^{\text{(b)}}\\\vdots\\\mathbf{E}_N^{\text{(b)}}\end{pmatrix},\\
    \begin{pmatrix}\mathbf{E}_1\\\vdots\\\mathbf{E}_N\end{pmatrix}&=
    \mathds{O}\begin{pmatrix}\mathbf{p}_1^\text{(fl)}\\\vdots\\\mathbf{p}_N^\text{(fl)}\end{pmatrix}+
    \mathds{P}\begin{pmatrix}\mathbf{E}_1^{\text{(b)}}\\\vdots\\\mathbf{E}_N^{\text{(b)}}\end{pmatrix},
  \end{split}
\end{equation}
and calculate the absorbed power \eqref{AbsP}. In the following $a_{i,\alpha}$ denotes the cartesian component $\bigl[\mathbf{a}_i\bigr]_\alpha$ ($\alpha=1,2,3$ corresponding to $x,y,z$ respectively) of the vector $\mathbf{a}_i$, whereas $\mathds{A}_{ij,\alpha\beta}$ the element $\bigl[\mathds{A}_{ij}\bigr]_{\alpha\beta}$ of the $3\times3$ matrix $\mathds{A}_{ij}$. From now on Latin indexes are associated to the dipoles, while Greek letters are used for cartesian components.  We have
\begin{align}
  p_{i,\alpha} &= \sum_{j,\beta}\Bigl(\mathds{M}_{ij,\alpha\beta}p_{j,\beta}^\text{(fl)}+\mathds{N}_{ij,\alpha\beta}E_{j,\beta}^\text{(fl)}\Bigr), \\
  E_{i,\alpha} &=\sum_{j,\beta}\Bigl(\mathds{O}_{ij,\alpha\beta}p_{j,\beta}^\text{(fl)}+\mathds{P}_{ij,\alpha\beta}E_{j,\beta}^\text{(fl)}\Bigr).
\end{align}

Now, assuming no correlation between the fluctuating dipole moments and the field of bath [i.e.\ $\langle{p}_{i,\alpha}^\text{(fl)}(\omega)E_{j,\beta}^{\text{(b)\dag}}(\omega')\rangle=0$ for any $\omega$ and $\omega'$, $i,j=1,\dots,N$ and $\alpha,\beta=x,y,z$] we get
\begin{equation}
  \label{pE}
  \begin{split}
    \langle\mathbf{p}_i(\omega)\cdot&\mathbf{E}_i^\dag(\omega')\rangle\\
         &=\sum_\alpha\sum_{jj'}\sum_{\beta\beta'}\Bigl[\mathds{M}_{ij,\alpha\beta}\langle p_{j,\beta}^\text{(fl)}p_{j',\beta'}^{'\text{(fl)\dag}}\rangle\mathds{O}^{'\dag}_{j'i,\beta'\alpha}\\
         &\quad+\mathds{N}_{ij,\alpha\beta}\langle E_{j,\beta}^\text{(fl)}E_{j',\beta'}^{'\text{(fl)\dag}}\rangle\mathds{P}^{'\dag}_{j'i,\beta'\alpha}\Bigr].
   \end{split}
\end{equation}
On the RHS the prime is associated to quantities calculated in $\omega'$, the others being calculated in $\omega$. The correlation functions appearing in Eq.~\eqref{pE} can be deduced from the fluctuation-dissipation theorem and read
\begin{equation}
  \label{FDTheorempP}
  \begin{split}
    \langle p_{j,\beta}^\text{(fl)}(\omega)p_{j',\beta'}^\text{(fl)\dag}(\omega')\rangle &=\hbar\,\varepsilon_0\delta_{jj'}\delta_{\beta\beta'}\chi_j2\pi\delta(\omega-\omega')\\
         & \quad\times\bigl[1+2n(\omega,T_j)\bigr] \\
   \end{split}
\end{equation}
and
\begin{equation}
  \begin{split}
  \label{FDTheorempE}
    \!\!\!\langle E^{\text{(b)}}_{j,\beta}(\omega)E^{\text{(b)}\dag}_{j',\beta'}(\omega')\rangle &=\frac{\hbar k^2}{\varepsilon_0}\Im(\mathds{G}_{jj',\beta\beta'}^{(0)})2\pi\delta(\omega-\omega')\\
         & \quad\times\bigl[1+2n(\omega,T_\text{b})\bigr].
  \end{split}
\end{equation}
Here we have introduced
\begin{equation}
  \chi_j =\Im(\alpha_j)-\frac{k^3}{6\pi}|\alpha_j|^2
\end{equation}
and  the Bose-Einstein distribution
\begin{equation}
   n(\omega,T) = \biggl[\exp\biggl(\frac{\hbar\omega}{\kb T}\biggr) - 1\biggr]^{-1}
\end{equation}
at temperature $T$. A discussion concerning the use of $\chi_j$ instead of $\Im(\alpha_j)$ is provided in Appendix \ref{AppDipole}.
By means of the expression for the correlations functions we conclude that
\begin{equation}
  \begin{split}
    \langle\mathbf{p}_i&(\omega)\cdot\mathbf{E}_i^\dag(\omega')\rangle \\
        &= 2\pi\delta(\omega-\omega')\\
        &\,\times\Bigl[\hbar\varepsilon_0\sum_j\chi_j\bigl[1+2n(\omega,T_j)\bigr]\Tr\Bigl(\mathds{M}_{ij}\mathds{O}^{\dag}_{ji}\Bigr)\\
        &\qquad+\frac{\hbar k^2}{\varepsilon_0}\bigl[1+2n(\omega,T_\text{b})\bigr]\Tr\Bigl(\mathds{N}\Im(\mathds{G}^{(0)})\mathds{P}^{\dag}\Bigr)_{ii}\Bigr].
  \end{split}
\end{equation}
Using Eqs.~\eqref{EqMatrixDip} and \eqref{EqMatrixE} we obtain
\begin{widetext}
\begin{equation}
  \label{pE2}
  \begin{split}
    \langle\mathbf{p}_i(\omega)\cdot\mathbf{E}_i^\dag(\omega')\rangle =
         2\pi\delta(\omega-\omega')&\Bigl[\hbar\varepsilon_0\sum_j\chi_j\bigl[1+2n(\omega,T_j)\bigr]\Bigl(\frac{1}{\varepsilon_0\alpha_i^*}+\frac{k^2}{\varepsilon_0}g_{ii}^{(0)*}\Bigr)\Tr\Bigl(\mathds{T}^{-1}_{ij}\mathds{T}^{-1\dag}_{ji}\Bigr)\\
     &\,-\hbar\varepsilon_0\chi_i\bigl[1+2n(\omega,T_i)\bigr]\frac{1}{\varepsilon_0\alpha_i^*}\Tr\Bigl(\mathds{T}^{-1}_{ii}\Bigr)\\
     &\,+\frac{\hbar k^2}{\varepsilon_0}\bigl[1+2n(\omega,T_\text{b})\bigr]\Bigl(\frac{1}{\varepsilon_0\alpha_i^*}+\frac{k^2}{\varepsilon_0}g_{ii}^{(0)*}\Bigr)\Tr\Bigl(\mathds{T}^{-1}\mathds{A}\Im(\mathds{G}^{(0)})\mathds{A}^{\dag}\mathds{T}^{-1\dag}\Bigr)_{ii}\Bigr],
  \end{split}
\end{equation}
where we have defined
\begin{equation}\mathds{G}_{ii}^{(0)}=g_{ii}^{(0)}\mathds{1}=\Bigl(a+i\frac{\omega}{6\pi c}\Bigr)\mathds{1},\qquad a\in\mathbb{R},\end{equation}
and introduced the (formally infinite) real part $a$ of the diagonal Green function $\mathds{G}_{ii}^{(0)}$ which will not play any role in the final results. Using the fact the the exponential factor $\re^{\ri(\omega-\omega')t}$ in Eq.~\eqref{AbsP} becomes irrelevant with respect to the imaginary part because of the delta function $\delta(\omega-\omega')$ appearing in Eq.~\eqref{pE2}, we obtain after simple algebraic manipulations
\begin{equation}
  \label{ImpE}
  \begin{split}
     \Im\langle\mathbf{p}_i(\omega)\cdot\mathbf{E}_i^\dag(\omega')\rangle&=2\pi\delta(\omega-\omega')\frac{\hbar\chi_i}{|\alpha_i|^2}\Bigl[\sum_j\chi_j\bigl[1+2n(\omega,T_j]\bigr)\Tr\Bigl(\mathds{T}^{-1}_{ij}\mathds{T}^{-1\dag}_{ji}\Bigr)-\bigl(1+2n(\omega,T_i)\bigr)\Im\Bigl[\alpha_i\Tr\Bigl(\mathds{T}^{-1}_{ii}\Bigr)\Bigr]\\
    &\quad+k^2\bigl[1+2n(\omega,T_\text{b})\bigr]\sum_{jk}\alpha_j\alpha_k^*\Tr\Bigl(\mathds{T}_{ij}^{-1}\Im(\mathds{G}_{jk}^{(0)})\mathds{T}_{ki}^{-1\dag}\Bigr)\Bigr].
   \end{split}
\end{equation}
\end{widetext}

It is physically evident that the net power absorbed by any dipole $i$ must be zero at thermal equilibrium. As a consequence, the following condition must hold for $i=1,\dots,N$
\begin{equation}\label{Cond0}\begin{split}\Tr&\Bigl[\sum_j\chi_j\mathds{T}^{-1}_{ij}\mathds{T}^{-1\dag}_{ji}-\Im\Bigl(\alpha_i\mathds{T}^{-1}_{ii}\Bigr)\\
&\,+k^2\sum_{jk}\alpha_j\alpha_k^*\mathds{T}_{ij}^{-1}\Im(\mathds{G}_{jk}^{(0)})\mathds{T}_{ki}^{-1\dag}\Bigr]=0.\end{split}\end{equation}
In Appendix \ref{AppN}, we discuss the cases of one and two dipoles, showing analytically that the condition \eqref{Cond0} is met. Furthermore, we have verified its validity for several higher values of $N$ and for random realizations of the geometrical configuration of the particles.

This condition allows us to write the net heat transfer on particle $i$ as a sum of exchanges with the other particles and with the bath
\begin{equation}\label{AbsPF}\begin{split}&\wp_i^\text{(abs)}(t,T_1,\dots,T_N,T_\text{b})\\
&=\int_0^{+\infty}\frac{d\omega}{2\pi}\hbar\omega\Bigl[\sum_{j\neq i}\frac{4\chi_i\chi_j}{|\alpha_i|^2}n_{ji}(\omega)\Tr\Bigl(\mathds{T}^{-1}_{ij}\mathds{T}^{-1\dag}_{ji}\Bigr)\\
&\,+\frac{4k^2\chi_i}{|\alpha_i|^2}n_{\text{b}i}(\omega)\sum_{jk}\alpha_j\alpha_k^*\Tr\Bigl(\mathds{T}_{ij}^{-1}\Im(\mathds{G}_{jk}^{(0)})\mathds{T}_{ki}^{-1\dag}\Bigr)\Bigr],\end{split}\end{equation}
where we have introduced the differences
\begin{equation}n_{ij}(\omega)=n(\omega,T_i)-n(\omega,T_j).\end{equation}

Equation \eqref{AbsPF} is one of the main results of this paper. It provides the expression of the instantaneous power absorbed by any dipole $i$ formally written as a sum of contributions associated to each other dipole $j$ and the thermal bath. It contains the polarizabilities of the $N$ dipoles through the terms $\alpha_i$ and $\chi_i$ and on the dependence on the geometrical configuration through the matrices $\mathbb{T}$ and $\mathbb{G}^{(0)}$. In the following we provide some numerical applications of this formula to the case of three dipoles.

\section{Numerical results}\label{SecNumerics}

In this section we present several numerical applications of the main formula \eqref{AbsPF}, applied to both time-dependent and time-independent configurations, in order to explore near-field many-body effects. We discuss here the case of three dipoles, the simplest one in which entangled interactions exist and where the heat transfer is not additive. We first study some examples of time evolution of the temperature distribution by varying the distance between the particle and thus show the role played by near-field interactions. Then, we discuss the importance of surface resonances and their coupling by varying the material properties of one of the three particles. Finally, we consider a time-independent case, and show that the ability of controlling the temperature of one of the particles combined with the geometrical distribution of the three dipoles can be exploited to tune the two other temperatures.

\subsection{Near-field heat exchange in a three-body system}

In this section we consider three identical spherical nano-particles having radii $R_i=50\,$nm ($i=1,2,3$) and made of silicon carbide (SiC). For the dielectric response of SiC we use the simple model~\cite{Palik98}
\begin{equation}
  \varepsilon(\omega) = \varepsilon_\infty\frac{\omega^2-\omega_{\rm l}^2+\ri\Gamma\omega}{\omega^2-\omega_{\rm t}^2+\ri\Gamma\omega},
\end{equation}
where $\varepsilon_\infty=6.7$, $\omega_{\rm l}~=~1.827\cdot10^{14}\,\mathrm{rad}\,\text{s}^{-1}$, $\omega_{\rm t}~=~1.495\cdot10^{14}\,\mathrm{rad}\,\text{s}^{-1}$ and $\Gamma=0.9\cdot10^{12}\,\mathrm{rad}\,\text{s}^{-1}$. This model implies a surface phonon-polariton resonance at $\omega_p=1.787\cdot10^{14}\,\mathrm{rad}\,\text{s}^{-1}$. For each particle $i$, we define the Clausius-Mossotti polarizability as
\begin{equation}
  \alpha_i^{(0)}(\omega) = 4\pi R_i^3\frac{\varepsilon_i(\omega)-1}{\varepsilon_i(\omega)+2}.
\end{equation}
The (dressed) polarizability $\alpha_i(\omega)$ for each dipole is then obtained by applying the radiative correction, discussed for example in~\cite{CarminatiOptCommun06,AlbaladejoOptExpress10},
\begin{equation}
  \label{AlphaDressed}
    \alpha_i(\omega)=\frac{\alpha_i^{(0)}(\omega)}{1-\ri\frac{k^3}{6\pi}\alpha_i^{(0)}(\omega)}.
\end{equation}
We remark here that the use of the dressed polarizability \eqref{AlphaDressed} makes the quantity $\chi_j$ appearing in Eq.~\eqref{FDTheorempP} always positive, and thus the energy flux is always in the correct direction, i.e.\ from hotter to colder particles.

\begin{figure*}[htb]
  \includegraphics[height=9cm]{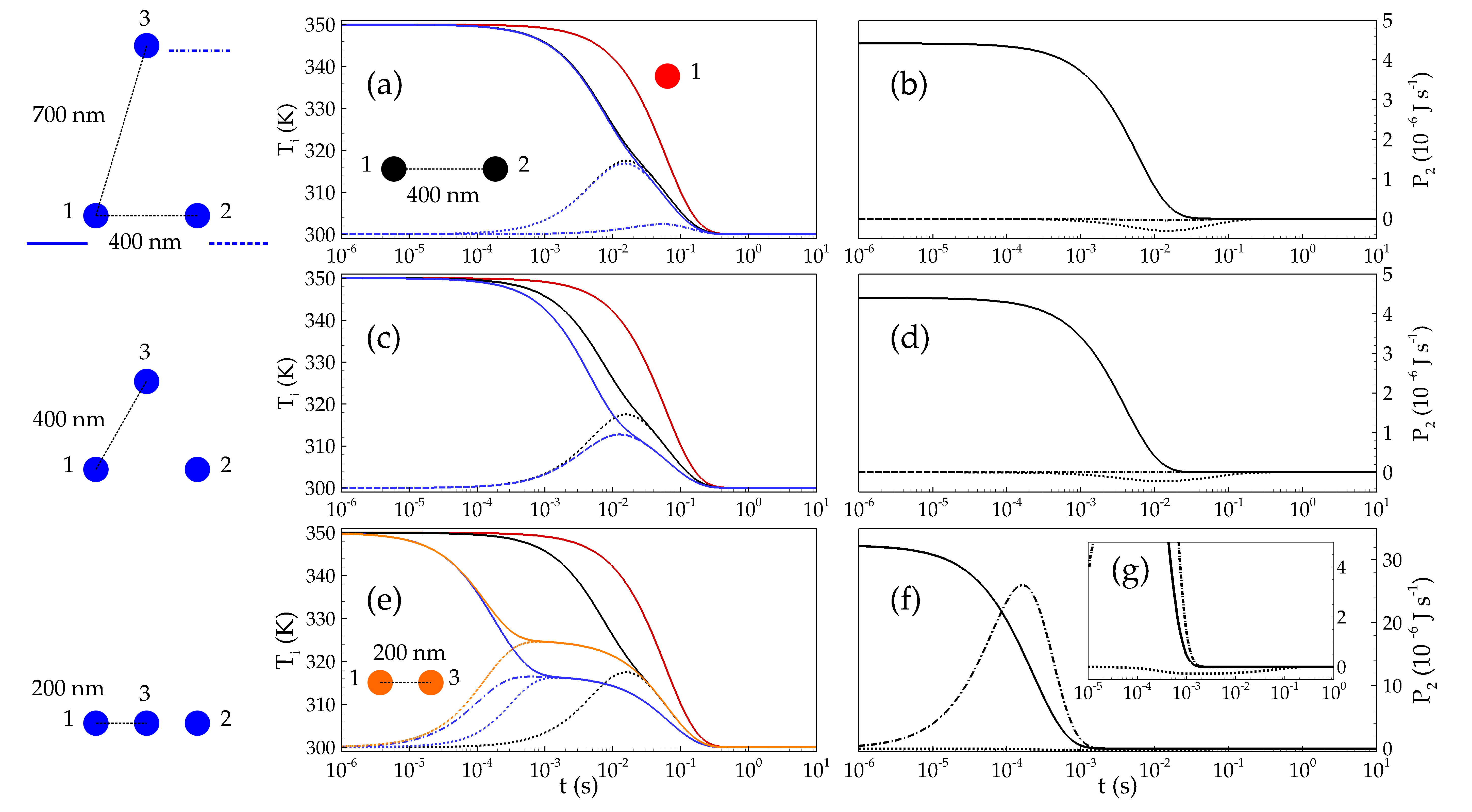}
  \caption{(color online) Panels (a), (c) and (e): time evolution of the temperatures in a three-body configuration. The distance between particles 1 and 2 is always 400\,nm, while the distances between dipole 1 (or 2) and 3 are (a) 700\,nm, (c) 400\,nm and (e) 200\,nm. The blue lines correspond to the three-body configuration (solid line for dipole 1, dashed line for dipole 2, dot-dashed line for dipole 3). The black lines correspond to the two-body case (solid line for dipole 1, dashed line for dipole 2), while the red solid line correspond to dipole 1 alone. In panel (e), we also show the two-body dynamics associated to dipoles 1 and 3 at a distance of 200\,nm (orange solid line for dipole 1, orange dot-dashed line for dipole 3). Panels (b), (d), (f) and (g) describe the time dependence of the power absorbed by particle 2. The solid line is the contribution coming from dipole 1, the dot-dashed line the contribution coming from dipole 3, the dashed line the power absorbed from the bath.}
  \label{Figure_1}
\end{figure*}

For any geometrical configuration, we solve the system of three differential equations \eqref{TempEv}, where the absorbed power is calculated using Eqs.~\eqref{AbsPF}, \eqref{DefTA} and \eqref{DefG0}. In order to reduce the number of degrees of freedom we place the dipole 1 at the origin ($\mathbf{R}_1=\mathbf{0}$), and dipole 2 in position $\mathbf{R}_2=(0,0,z_2)$, at a distance of $z_2=400\,$nm (see Fig.~\ref{Figure_1}). As for dipole 3, we fix its $z$ coordinate as $z_3=z_2/2$ and vary its $y$ coordinate. In Fig.~\ref{Figure_1} we show three different configurations in which the distances between dipole 1 (or 2) and 3 are 700\,nm (panels (a)-(b)), 400\,nm (panels (c)-(d)) and 200\,nm (panels (e)-(g)). For each geometry, we study the time evolution of the three temperatures with the initial conditions $(T_1(0),T_2(0),T_3(0))=(350,300,300)\,$K and $T_\text{b}=300\,$K. We thus assume that from a configuration in which the entire system was at thermal equilibrium at ambient temperature, we heat one of the particles (dipole 1) up to 350\,K. The evolution of the three temperatures is compared to the evolution of one single dipole heated up to 350\,K (red curve in figure) and to the case of two dipoles (1 and 2) at a distance of 400\,nm (black curves in figure). Our interest is in particular to show how the presence of a third particle modifies the thermalization process of particle 2.

As expected on physical grounds, Fig.~\ref{Figure_1} shows that the three dipoles thermalize to the temperature of the bath $T_\text{b}=300\,$K. For our choices of materials and distances, this process takes approximately 1\,s, in presence of one, two or three dipoles and is apparently almost independent from the geometrical configuration. On the contrary, it is manifest that a different time scale exists associated to a thermalization process taking place between the three particles. In the first case [see Fig.~\ref{Figure_1}(a)], the distance between dipoles 1 (or 2) and 3 is such that the presence of dipole 3 plays a negligible role on the dynamics of the temperatures of dipoles 1 and 2, which is very close to the two-body case. In this case, dipoles 1 and 2 thermalize between each other around $t=10^{-2}\,$s, whereas the temperature of dipole 3 is modified very weakly and locally in time with respect to the equilibrium value of 300\,K. The situation is clearly different in the case depicted in Fig.~\ref{Figure_1}(c), corresponding to an equilateral triangle. In this case, as obvious from symmetry arguments, dipoles 2 and 3 follow exactly the same evolution and the figure shows the existence (as in the two-dipole case) of a different time scale associated to near-field interactions. The third and last case [Fig.~\ref{Figure_1}(e)], in which the dipoles are aligned and the minimum distance is 200\,nm instead of 400\,nm, proves first of all that this new time scale is extremely sensitive to the distance between the dipoles. In this case, dipole 3 is heated faster than dipole 2, coherently with the fact that it is closer to dipole 1. Nevertheless, it is clear that reducing the distance between dipole 3 and 2 as well makes dipole 3 act as a bridge between dipoles 1 and 2 producing a remarkable acceleration (of approximately one order of magnitude) of its temperature dynamics. In this last case, we can clearly identify an interval of time during which thermalization between dipoles is produced, at a temperature significantly different from the one of the bath. In this case, we also compare the three-body result to the two-body case at a distance of 200\,nm. We clearly identify an interval of time during which dipole 2 has a temperature still close to 300\,K, while the temperature of dipole 3 deviates from 300\,K following the two-body dynamics. This clearly proves that the time scale associated to the dynamics at a distance of 200\,nm is significantly faster than the one corresponding to $d=400\,$nm.

From this numerical example it becomes apparent that the smallest distance between particles determines the time scale on which the heat flux is exchanged between the particles. Further, for the distances considered in the present work this times scale is still several orders of magnitude larger than the one associated with internal phonon thermalization inside each dipole. This justifies the assumption made at the beginning (see Sec.~\ref{SecPhysSyst}) allowing us to associate a temperature to each particle as a function of time.

Some more insight on the temperature dynamics is given by panels (b), (d), (f) and (g) of Fig.~\ref{Figure_1}, where the power absorbed by dipole 2 is represented for the three cases under scrutiny. This power is decomposed in the three contributions coming from dipole 1, dipole 3 and from the bath. For small $t$, the power absorbed by dipole 2 comes almost entirely from dipole 1, as expected. Moreover, in panel (b) the distance between dipoles 3 and 2 is such that the power exchanged between them is negligible, while around $t=10^{-2}\,$s the temperature difference between dipoles 1 and 2 is such that the (negative) power absorbed by dipole 2 and coming from the bath starts being comparable (and later on larger) to the exchange between dipoles 1 and 2. This comparison shows that, even in the near field, at some point the temperature difference and the intra-dipole thermalization fixes the time interval during which only the far-field exchange with the bath matters. As far as panel (d) is concerned, no power is exchanged between dipoles 3 and 2, since their temperatures always coincide. Nevertheless, it is interesting to underline that in this case the power exchanged with the bath is (slightly) modified with respect to the first case. This proves that even the far-field interaction is affected by the geometrical configuration and near-field properties. The third case (with the three dipoles aligned) has a dramatically different power-time diagram with respect to the first two cases. In this case, after a strong exchange with dipole 1, dipole 2 starts absorbing more energy from dipole 3, which is hotter than dipole 2 [see Fig.~\ref{Figure_1}(e)]. We also see that intra-dipole power exchanges become negligible around $t=10^{-3}\,$s, time at which the thermalization between the particles has almost finished. Finally, panel (g) shows, in the same power scale of the previous ones, that the exchange with the bath is again modified by near-field properties, and in particular accelerated by about one order of magnitude.

In the next section, we will see how changing the material properties of one of the three dipoles affects near-field effects and temperature dynamics.

\subsection{Dependence of dynamic relaxation on surface resonances}

In the previous section, the three particles have always been considered to be made of the same material (SiC). It is well known that even in stationary configuration this choice maximizes the heat flux, since it produces the best possible coupling between surface modes (phonon-polaritons, in the considered case of a polar material), which give the main contribution to heat transfer in near field~\cite{JoulainSurfSciRep05}. In order to see how the dynamics changes if this coupling is no longer present, we consider a specific geometrical configuration, and analyze the cases in which one particle at a time is replaced with a different material.

\begin{figure*}[htb]
\includegraphics[height=7cm]{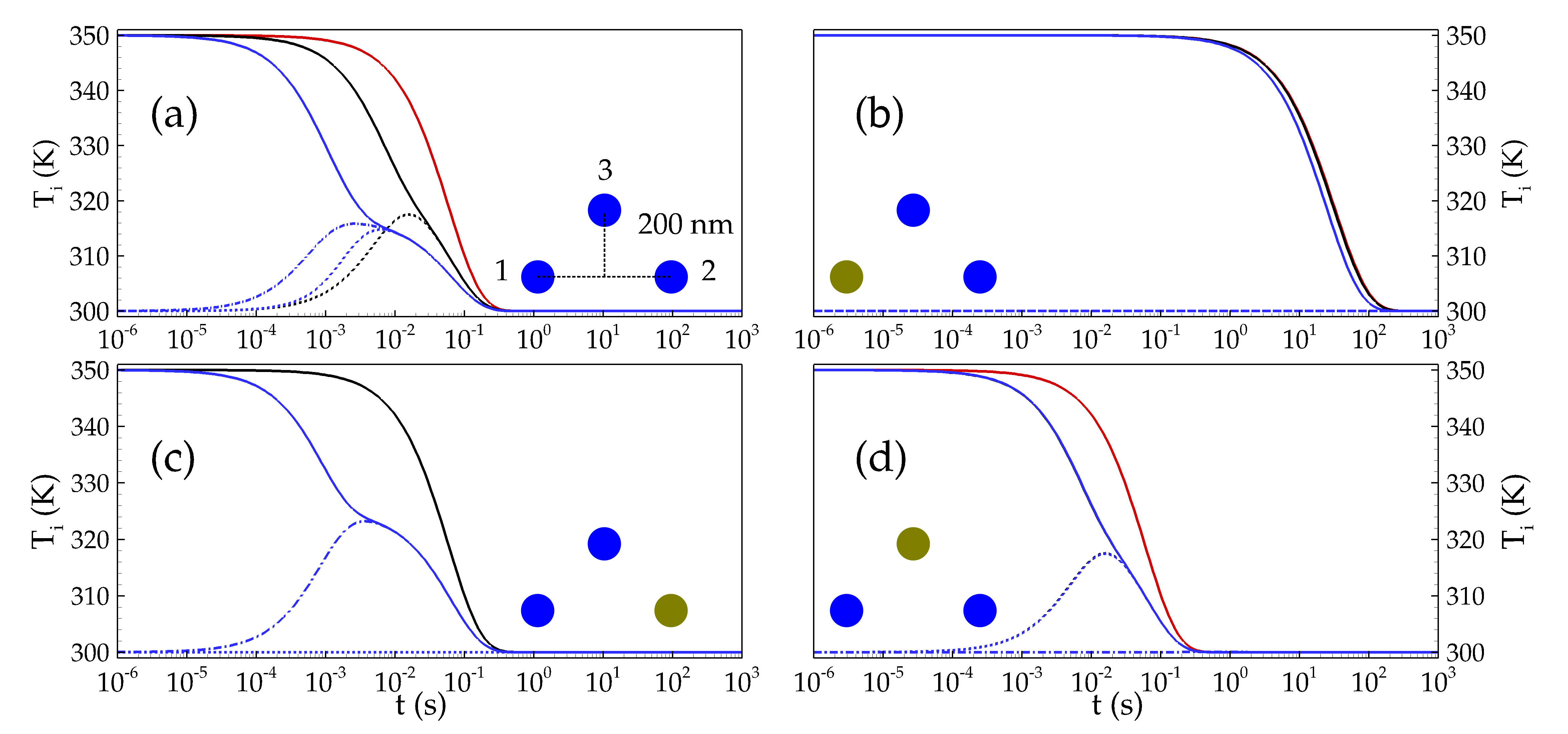}
  \caption{(color online) Time evolution of the temperatures in a three-body configuration. Same color convention of Fig.~\ref{Figure_1}. Panel (a): three SiC particles. In panels (b), (c) and (d) particles 1, 2 and 3 are respectively replaced by a gold nanosphere. We remark that in (b) the red and black curves relative to particle 1 are almost superposed. The same is true in (c), while in (d) the blue and black curves relative to particles 1 and 2 are superposed.}
  \label{Figure_2}
\end{figure*}

To be more specific, we consider a set of coordinates $\mathbf{R}_1=\mathbf{0}$, $\mathbf{R}_2=(0,0,400)\,$nm, and $\mathbf{R}_2=(0,200,200)\,$nm. Two among the three particles are made of SiC, while the third one is made of gold, described using a Drude model
\begin{equation}
  \varepsilon(\omega)=1-\frac{\omega_\text{pl}^2}{\omega(\omega+ {\rm i}\gamma)}
\end{equation}
with $\omega_\text{pl}=1.37\cdot10^{16}\,\mathrm{rad}\,\text{s}^{-1}$ and $\gamma~=~0.4~\cdot~10^{14}\,\mathrm{rad}\,\text{s}^{-1}$. The three cases in which one of the SiC particles is replaced by a gold one are compared in Fig.~\ref{Figure_2} to the case of three SiC particles. The choice of gold is motivated by the fact that the plasmon resonance it supports lies in the ultraviolet range, thus both far from the resonance of SiC and outside the region where the population $n(\omega,T)$ takes non-negligible values at the chosen temperatures.

In Fig.~\ref{Figure_2}(a) the standard case of three SiC particles is represented. We see the effects already discussed in the previous Section, and in particular the possibility of modifying the time scale of thermalization thanks to near-field interactions. Figure \ref{Figure_2}(b) represents the case in which particle 1, the only particle heated up to 350\,K in the system, is made of gold. We observe two phenomena: first of all the coupling between particle 1 and particles 2 and 3 is almost absent, and the temperatures of both particles 2 and 3 remain close to 300\,K during the entire process. Moreover, also the time scale of thermalization toward the temperature of the bath is modified. This modification is due to the fact, anticipated before, that the resonance of gold is at a frequency at which the population $n(\omega,T)$ is negligible, and as a consequence the coupling (even with the bath) is much weaker with respect to the case of SiC. In Fig.~\ref{Figure_2}(c) the second particle is replaced with a gold one. In this case we see that its temperature is at any time indistinguishable from 300\,K, meaning that it does not feel any coupling to particles 1 and 3. For these particles, we observe, on the contrary, a typical two-body dynamics, with a thermalization between the dipoles taking place more quickly than the one toward the bath temperature. In Fig.~\ref{Figure_2}(d), finally, we observe a two-body dynamics between particles 1 and 2 [the same described by black curves in Fig.~\ref{Figure_2}(a)], while particle 3, made of gold, does not participate to the energy exchange. Figures \ref{Figure_2}(c) and \ref{Figure_2}(d) present indeed two different temperature dynamics, the difference being the fact the particles 1 and 3 are closer than particles 1 and 2.

\subsection{Steady state and temperature control in a three-body system}

In this section we focus our attention on a stationary problem, that is the distribution of temperatures among the particles for $t\to+\infty$. In this limit the LHS of Eq.~\eqref{TempEv} is zero such that $\wp_i^\text{(abs)}$ is zero for all particles. It is evident that, for any choice of initial temperatures $T_i(0)$, without an external source of energy the temperatures in the long-time limit coincide with $T_\text{b}$. We thus assume in this section that one of the particles, say particle 1, is heated up to 350\,K as in the time-dependent simulations, but kept at this temperature by means of a thermostat. We are interested in showing how the positions of particles 2 and 3 modify the temperatures these particle assume for $t\to+\infty$. To this aim, for a given geometrical configuration we calculate the matrices $\mathbb{T}$ and $\mathbb{G}^{(0)}$, the power absorbed by particles 2 and 3 [using Eq.~\eqref{AbsPF}], and impose that these powers are zero in order to find $T_2$ and $T_3$.

\begin{center}\begin{figure}[htb]
\includegraphics[height=7cm]{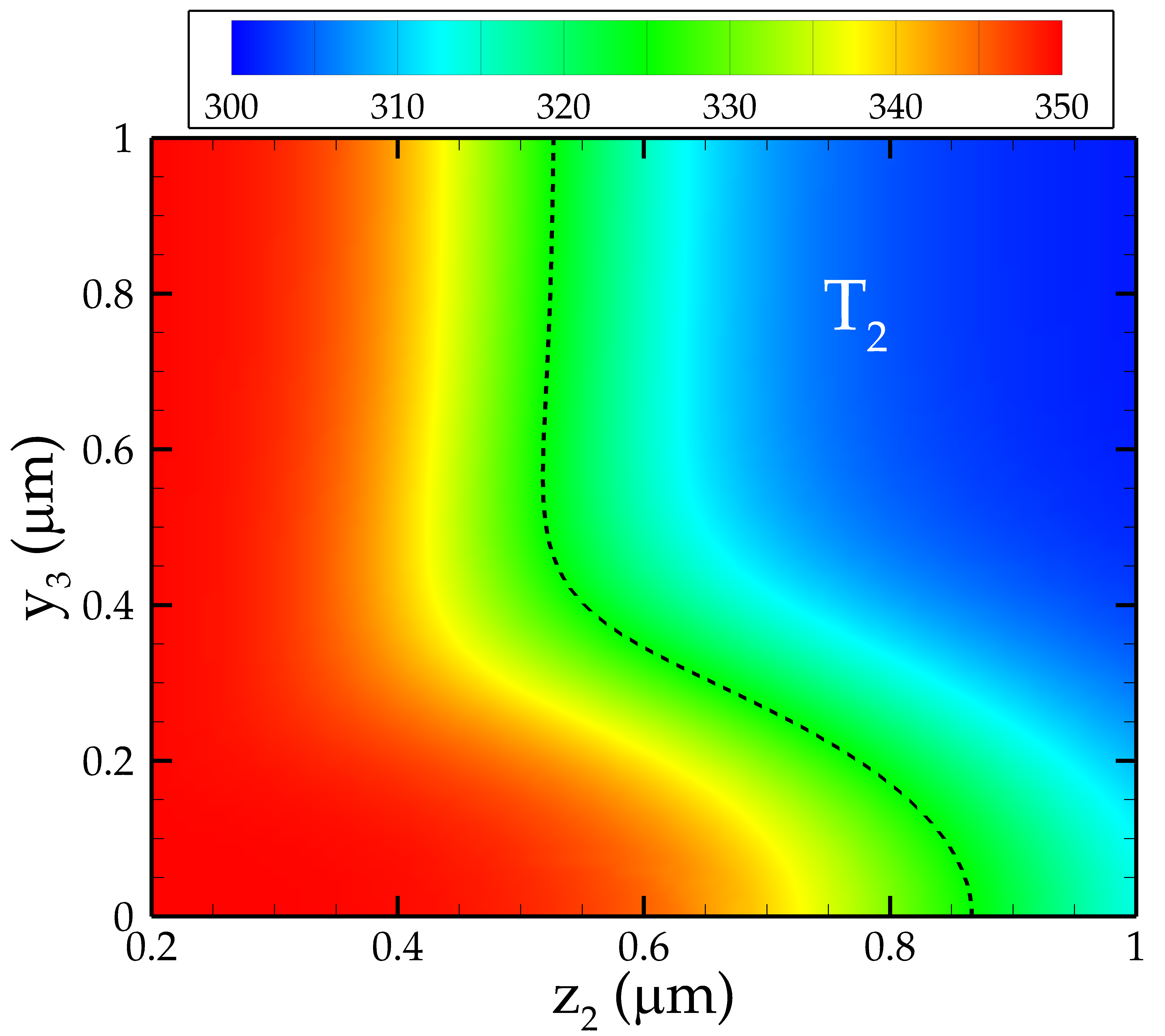}
\caption{(color online) Equilibrium temperature of particle 2 when the three particles have coordinates $\mathbf{R}_1=(0,0,0)$, $\mathbf{R}_2=(0,0,z_2)$ and $\mathbf{R}_3=(0,y_3,z_2/2)$. Particle 1 is kept at temperature $T_1=350\,$K and the bath has temperature $T_\text{b}=300\,$K. The black dashed line corresponds to $T_2=325\,$K.}\label{Figure_3}\end{figure}\end{center}

In order to reduce the number of degrees of freedom we consider the case in which particle 1 is placed at the origin, particle 2 has coordinates $\mathbf{R}_2=(0,0,z_2)$ and particle 3 is located in $\mathbf{R}_3=(0,y_3,z_2/2)$. We are left with two independent variables, namely $z_2$ and $y_3$, as a function of which we study the equilibrium temperature $T_2$ of particle 2. The result is shown in Fig.~\ref{Figure_3}, where $z_2$ varies in the range $[200\,\text{nm},1\,\mu\text{m}]$ and $y_3$ in $[0,1]\,\mu\text{m}$.

We immediately notice that for any considered geometry the temperature $T_2$ lies, as expected, in the range $[300,350]\,$K, i.e.\ between the temperatures of particle 1 and of the bath. Moreover, we see that starting from values of $y_3$ of the order of 500\,nm the presence of particle 3 no longer plays a role: on a given horizontal line we are left with the distribution of equilibrium temperature particle 2 would have in presence of particle 1 only. As expected, this distribution goes to 300\,K (to $T_\text{b}$) when $z_2$ increases, and in particular for $z_2$ above 700\,nm. On the contrary, $T_2$ is close to 350\,K (to $T_1$) for $z_2$ smaller than 300\,nm. When the coordinate $y_3$ of particle 3 is modified, the dependence of $T_2$ on $z_2$ is deeply affected, and in particular for small $y_3$, between 0 and 200\,nm, even for $z_2$ as large as 1\,$\mu$m the temperature $T_2$ is still close to the average between $T_1$ and $T_\text{b}$. This gives an alternative evidence of the fact that the presence of particle 3 can act as a bridge for near-field interaction between the external particles 1 and 2. Moreover, this calculation shows that a localized heating and the use of few external energy sources can be actively exploited to produce a desired time-independent temperature profile in a collection of dipoles by acting on their geometrical distribution.

\section{Conclusions}\label{SecConcl}

We have used a purely fluctuational-electrodynamic approach to deduce the power absorbed by each particle in a collection of $N$ particles described as $N$ dipoles immersed in a thermal bath. These powers have been used to study the time evolution of $N$ temperatures with respect to different initial conditions. We have also addressed the study of the distribution of temperatures when one of the particle temperatures is kept fixed in time by applying a thermostat.

First of all, we have shown that near-field interactions introduce a different time scale of thermalization compared with the one associated wth far-field exchanges with the thermal bath. At short distances, in the regime of near-field interaction (typically for distances of the order of 100\,nm), the system shows first a thermalization between the particles, which then behave as a complex system thermalizing towards the bath temperature. The difference between these two time scales can go up to approximately two orders of magnitude by tuning the interaction between the nanoparticles. We have shown numerically that the intra-particle relaxation is extremely sensitive to the distance and we have also shown that, even in the simple case of three particles, the third particle modifies the temperature dynamics of the two others and also the time-dependent power they exchange between each other. We have also proved that this phenomenon depends strongly on the existence and the frequency of surface resonances: the coupling decreases drastically if the particles do not share a common surface mode. Finally, we have also considered the case in which the temperature of one of the particles is fixed in time, showing that the positions of the other particles can be used to manipulate their equilibrium temperature.

Our results show that many-body near-field interactions constitute a promising tool to tailor both time-dependent and time-independent heat fluxes and temperature distributions in a complex plasmonic system. This work paves the way to several interesting developments. First of all, it will be interesting to understand how these phenomena depend on the number of particles, by understanding whether collective phenomena can occur. Furthermore, heat spreading can be studied using our formalism, in order to see whether anomalous propagation regimes are possible or not because of the presence of $N$-body interactions. Finally, the problem of how multipolar contributions influence the many-body coupling has to be addressed as well.

\begin{acknowledgments}
The authors thank F. J. Garc\'{\i}a de Abajo, G. Dedkov, C. Henkel, K. Joulain, and A. I. Volokitin for fruitful discussions. P. B.-A. acknowledges the support of the Agence Nationale de la Recherche through the Source-TPV project ANR 2010 BLANC 0928 01. M. T. gratefully acknowledges support from the Stiftung der Metallindustrie im Nord-Westen.
\end{acknowledgments}

\appendix

\section{Fluctuation-dissipation theorem for a dipole}\label{AppDipole}

In this section we discuss the derivation of the correlation functions of a fluctuating dipole at temperature $T$ given in Eq.~\eqref{FDTheorempP}. In particular, we justify the use of $\chi_j=\Im(\alpha_j)-\frac{k^3}{6\pi}|\alpha_j|^2$ instead of the simpler factor $\Im(\alpha_j)$ typically used in literature. The quantity $\chi_j$ was already recently used in~\cite{ManjavacasPRB12}, without providing a detailed derivation. The derivation of the correlation functions $\langle p_{j,\beta}^\text{(fl)}(\omega)p_{j',\beta'}^\text{(fl)\dag}(\omega')\rangle$ appearing in Eq.~\eqref{FDTheorempP} starts from the calculation of the correlation functions of the electric field emitted by the fluctuating dipole, demanding a careful use of the fluctuation-dissipation theorem. The assumption of having bodies at fixed different temperatures out of thermal equilibrium is usually refereed to as local thermal equilibrium. Starting from the pioneering work of Polder and van Hove~\cite{PolderPRB71} and Rytov~\cite{Rytov89} this hypothesis is considered as equivalent to the statement that the field emitted by each body has the same statistical properties it would have if the body under scrutiny was at thermal equilibrium at its temperature. This issue is discussed in detail, for example, in several works presenting general theories for Casimir force and heat transfer out of thermal equilibrium~\cite{BimontePRA09,MessinaEurophysLett11,MessinaPRA11,KrugerPRL11}.

Let us then consider a fluctuating dipole at temperature $T$ at thermal equilibrium. The definition of thermal equilibrium implies that the dipole \emph{must} be immersed in a bath at the same temperature in such a way that the power radiated by the dipole equals the one absorbed from the bath. For this system the total field in any point of space is the sum of the one emitted by the dipole, the one coming from the bath, and the one scattered by the dipole. Being at thermal equilibrium, the fluctuation-dissipation theorem can be directly applied to the total field. The correlation functions of the field coming from the bath, described as a free bosonic field, are known. Since the connection between the induced dipole and the external field is established [see Eq.~\eqref{IndDip}], the scattered field is known as well. Finally, the correlation functions of the emitted field can be deduced. This procedure is described in detail in~\cite{MessinaPRA11} for an arbitrary body (not necessarily in the dipolar approximation).

The field emitted by the fluctuating dipole can then be written as
\begin{equation}
   \mathbf{E}^{\text{(dip)}}(\mathbf{R},\omega)=\frac{1}{4\pi\epsilon_0}\nabla_\mathbf{R}\times\nabla_\mathbf{R}\times\Biggl[\mathbf{p}^{\text{(fl)}}\frac{\re^{\ri\frac{\omega}{c}R_d}}{R_d}\Biggr]
\end{equation}
where $R_d=|\mathbf{R}_d|=|\mathbf{R}-\mathbf{R}_p|$ ($\mathbf{R}_p$ being the position of the dipole) and $\nabla_\mathbf{R}$ represents the gradient with respect to $\mathbf{R}$. The result previously obtained and this last equation finally allows us to prove directly the dipolar correlation functions appearing in Eq.~\eqref{FDTheorempP}.

\section{Cases of $N=1$ and $N=2$}\label{AppN}

We provide in this section the explicit expression of the $\mathds{T}$ in the case of one and two dipoles immersed in a thermal bath. For these two cases, we analytically show that the net absorbed power is zero at thermal equilibrium.

\subsection{One dipole}

In this case it follows immediately from Eq.~\eqref{DefTA} that
\begin{equation}
  \mathds{T}=\mathds{T}^{-1}=\mathds{1},\qquad\mathds{A}=\varepsilon_0\alpha_1\mathds{1},
\end{equation}
and then Eq.~\eqref{ImpE} becomes
\begin{equation}
  \begin{split}
    \Im&\langle\mathbf{p}_i(\omega)\cdot\mathbf{E}_i^\dag(\omega')\rangle \\
             &=2\pi\delta(\omega-\omega')\frac{\hbar\chi_1}{|\alpha_1|^2} \Bigl[\chi_1\bigl[1+2n(\omega,T_1)\bigr]\Tr(\mathds{1})\\
             &\qquad-\bigl[1+2n(\omega,T_1)\bigr]\Im(\alpha_1)\Tr(\mathds{1})\\
             &\qquad\,+k^2\bigl[1+2n(\omega,T_\text{b}]\bigr)|\alpha_1|^2\Tr\Bigl(\Im(\mathds{G}_{11}^{(0)})\Bigr)\Bigr].
  \end{split}
\end{equation}

From this equation we clearly see that no net power is exchanged between the dipole and the bath for $T_1=T_\text{b}$ and we finally deduce the simple formula
\begin{equation}\Im\langle\mathbf{p}_1(\omega)\cdot\mathbf{E}_1^\dag(\omega')\rangle=2\pi\delta(\omega-\omega')\frac{\hbar k^3}{\pi}\chi_1n_{b1}(\omega),\end{equation}
describing the spectral power density associated with the thermalization of a single dipole in a thermal bath.

\subsection{Two dipoles}

In this case we have

\vspace{3cm}
\begin{widetext}
\begin{alignat}{4}
     \mathds{T}&=\begin{pmatrix}
                   \mathds{1} & -k^2\alpha_1\mathds{G}_{12}^{(0)}
                   -k^2\alpha_2\mathds{G}_{21}^{(0)} & \mathds{1}
                  \end{pmatrix},
     \qquad\qquad
     &\mathds{T}^{-1} &=\begin{pmatrix}
                          \frac{1}{\mathds{P}} & k^2\alpha_1\frac{\mathds{G}_{12}^{(0)}}{\mathds{P}}\\
                          k^2\alpha_2\frac{\mathds{G}_{21}^{(0)}}{\mathds{P}} & \frac{1}{\mathds{P}}
                        \end{pmatrix}, \\
     \mathds{T}^{-1\dag} &=\begin{pmatrix}
                             \frac{1}{\mathds{P}^\dag} & k^2\alpha_2^*\frac{\mathds{G}_{21}^{(0)\dag}}{\mathds{P}^\dag}\\
                             k^2\alpha_1^*\frac{\mathds{G}_{12}^{(0)\dag}}{\mathds{P}\dag} & \frac{1}{\mathds{P}^\dag}
                           \end{pmatrix},
     \qquad\qquad \hfill
     &\mathds{P} &= \mathds{1}-k^4\alpha_1\alpha_2\mathds{G}_{12}^{(0)}\mathds{G}_{21}^{(0)}.
\end{alignat}
We now calculate Eq.~\eqref{ImpE} for $i=1$ (the case $i=2$ is equivalent). We have
\begin{equation}
  \begin{split}
    \Im\langle\mathbf{p}_1(\omega)\cdot\mathbf{E}_1^\dag(\omega')\rangle &= 2\pi\delta(\omega-\omega')\frac{\hbar\chi_1}{|\alpha_1|^2}\Bigl\{\chi_1\bigl[1+2n(\omega,T_1)\bigr]\Tr\Bigl(\mathds{T}^{-1}_{11}\mathds{T}^{-1\dag}_{11}\Bigr) + \chi_2\bigl[1+2n(\omega,T_2)\bigr]\Tr\Bigl(\mathds{T}^{-1}_{12}\mathds{T}^{-1\dag}_{21}\Bigr)\\
       &\quad-\bigl[1+2n(\omega,T_1)\bigr]\Im\Bigl[\alpha_1\Tr\Bigl(\mathds{T}^{-1}_{11}\Bigr)\Bigr] + k^2\bigl[1+2n(\omega,T_\text{b})\bigr] \Tr\Bigl[|\alpha_1|^2 \mathds{T}_{11}^{-1}\Im(\mathds{G}_{11}^{(0)})\mathds{T}_{11}^{-1\dag} \\
       &\quad + \alpha_1\alpha_2^* \mathds{T}_{11}^{-1}\Im(\mathds{G}_{12}^{(0)})\mathds{T}_{21}^{-1\dag} + \alpha_2\alpha_1^* \mathds{T}_{12}^{-1}\Im(\mathds{G}_{21}^{(0)})\mathds{T}_{11}^{-1\dag} + |\alpha_2|^2 \mathds{T}_{12}^{-1}\Im(\mathds{G}_{22}^{(0)})\mathds{T}_{21}^{-1\dag}\Bigr]\Bigr\},
   \end{split}
\end{equation}
and then
\begin{equation}
  \begin{split}
     \Im\langle\mathbf{p}_1(\omega)\cdot\mathbf{E}_1^\dag(\omega')\rangle
          &= 2\pi\delta(\omega-\omega')\frac{\hbar\chi_1}{|\alpha_1|^2}\Biggl\{\chi_1\bigl[1+2n(\omega,T_1)\bigr]\Tr\Bigl(\frac{1}{\mathds{P}\mathds{P}^{\dag}}\Bigr) + \chi_2\bigl[1+2n(\omega,T_2)\bigr]k^4|\alpha_1|^2\Tr\Bigl(\frac{\mathds{G}_{12}^{(0)}\mathds{G}_{12}^{(0)\dag}}{\mathds{P}\mathds{P}^{\dag}}\Bigr)\\
        &\quad -\bigl[1+2n(\omega,T_1)\bigr]\Im\Bigl[\alpha_1\Tr\Bigl(\frac{1}{\mathds{P}}\Bigr)\Bigr] + k^2\bigl[1+2n(\omega,T_\text{b})\bigr] \Tr \biggl( \frac{1}{\mathds{P}\mathds{P}^{\dag}}\Bigl[|\alpha_1|^2\frac{k}{6\pi} \mathds{1} \\
         &\quad     + \frac{k^5}{6\pi}|\alpha_1\alpha_2|^2 \mathds{G}_{12}^{(0)}\mathds{G}_{12}^{(0)\dag}
                     + k^2|\alpha_1|^2\alpha_2^*\Im(\mathds{G}_{12}^{(0)})\mathds{G}_{12}^{(0)\dag}
                     + k^2|\alpha_1|^2\alpha_2 \Im(\mathds{G}_{21}^{(0)})\mathds{G}_{12}^{(0)}\Bigr]\biggr)\Biggr\}.
  \end{split}
\end{equation}
We conclude that
\begin{equation}
  \begin{split}
    \Im\langle\mathbf{p}_1(\omega)\cdot\mathbf{E}_1^\dag(\omega')\rangle
        &= 2\pi\delta(\omega-\omega')\hbar\chi_1\Biggl\{\frac{\chi_1}{|\alpha_1^2|}\bigl[1+2n(\omega,T_1)\bigr]\Tr\Bigl(\frac{1}{\mathds{P}\mathds{P}^{\dag}}\Bigr)+k^4\chi_2\bigl[1+2n(\omega,T_2]\bigr)\Tr\Bigl(\frac{\mathds{G}_{12}^{(0)}\mathds{G}_{12}^{(0)\dag}}{\mathds{P}\mathds{P}^{\dag}}\Bigr)\\
       &\quad -\bigl[1+2n(\omega,T_1)\bigr]\Im\Bigl[\frac{1}{\alpha_1^*}\Tr\Bigl(\frac{1}{\mathds{P}}\Bigr)\Bigr] + k^2\bigl[1+2n(\omega,T_\text{b})\bigr] \Tr\biggl(\frac{1}{\mathds{P}\mathds{P}^{\dag}}\Bigl[\frac{k}{6\pi} \mathds{1}
         + \frac{k^5}{6\pi}|\alpha_2|^2 \mathds{G}_{12}^{(0)}\mathds{G}_{12}^{(0)\dag} \\
       &\quad + 2k^2 \Im(\mathds{G}_{21}^{(0)})\Re(\alpha_2\mathds{G}_{12}^{(0)})\Bigr]\biggr)\Biggr\}.
\end{split}\end{equation}
This quantity is zero at thermal equilibrium, and can thus be rewritten into the form of an exchange with particle 2 and with the bath
\begin{equation}
  \label{Eq:power21}
  \begin{split}
    \Im\langle\mathbf{p}_1(\omega)\cdot\mathbf{E}_1^\dag(\omega')\rangle
       &=2\pi\delta(\omega-\omega')\Bigl\{n_{21}(\omega)2\hbar k^4\chi_1\chi_2\Tr\Bigl(\frac{\mathds{G}_{12}^{(0)}\mathds{G}_{12}^{(0)\dag}}{\mathds{P}\mathds{P}^{\dag}}\Bigr)\\
       &\quad + n_{\text{b}1}(\omega)\frac{2\hbar k^3\chi_1}{6\pi}\Tr\biggl( \frac{1}{\mathds{P}\mathds{P}^{\dag}}\Bigl[\frac{1}{6\pi} \mathds{1} + \frac{k^4}{6\pi}|\alpha_2|^2\mathds{G}_{12}^{(0)}\mathds{G}_{12}^{(0)\dag} + 2k \Im\bigl(\mathds{G}_{12}^{(0)}\bigr)\Re\bigl(\alpha_2\mathds{G}_{12}^{(0)\dag}\bigr)\Bigr]\biggr)\Bigr\}.
  \end{split}
\end{equation}
\end{widetext}

\end{document}